\documentclass[manuscript]{aastex}
\usepackage[export]{adjustbox}
\usepackage{ragged2e}
\usepackage{enumitem}
\usepackage{lineno}
\shorttitle{Analyzing the Sequence of Phases Leading to the Formation of the Active Region 13664}
\shortauthors{Romano, Elmhamdi, Marassi, and Contarino}

\begin{document}
\justifying

%\title{Homologous flares associated with Sigmoidal structures in the new Solar Cycle Active Region NOAA 12790 }
\title{Analyzing the Sequence of Phases Leading to the Formation of the Active Region 13664, with Potential Carrington-like Characteristics}

%\title{A magnetically simple AR of the new SC25 hosting a sequence of peculiar C-class flares}

\author{P. Romano\altaffilmark{1}, A. Elmhamdi\altaffilmark{2}, A. Marassi\altaffilmark{3} and L. Contarino\altaffilmark{1}}

%\affil{Astronomy Department, University of California,
%    Berkeley, CA 94720}

%\author{C. D. Biemesderfer\altaffilmark{4,5}}
%\affil{National Optical Astronomy Observatories, Tucson, AZ 85719}
\email{paolo.romano@inaf.it}

%\and

%\author{R. J. Hanisch\altaffilmark{5}}
%\affil{Space Telescope Science Institute, Baltimore, MD 21218}

\altaffiltext{1}{INAF - Osservatorio Astrofisico di Catania,
              Via S. Sofia 78, 95123 Catania, Italy.}
\altaffiltext{2}{Department of Physics and Astronomy, King Saud University, 11451 Riyadh, Saudi Arabia.}
\altaffiltext{3}{INAF - Osservatorio Astronomico di Trieste, Via G.B. Tiepolo 11, 34143 Trieste, Italy.}

%\altaffiltext{3}{present address: Center for Astrophysics,
%    60 Garden Street, Cambridge, MA 02138}
%\altaffiltext{4}{Visiting Programmer, Space Telescope Science Institute}
%\altaffiltext{5}{Patron, Alonso's Bar and Grill}


\begin{abstract}
Several recurrent X-class flares from Active Region (AR) 13664 have triggered a severe G5-class geomagnetic storm between May 10 and 11, 2024. The morphology and compactness of this AR closely resemble the active region responsible for the famous Carrington Event of 1859. Although the induced geomagnetic currents produced a value of the Dst index, probably, an order of magnitude weaker than that of the Carrington Event, the characteristics of AR 13664 warrant special attention. Understanding the mechanisms of magnetic field emergence and transformation in the solar atmosphere that lead to the formation of such an extensive, compact and complex AR is crucial. Our analysis of the emerging flux and horizontal motions of the magnetic structures observed in the photosphere reveals the fundamental role of a sequence of emerging bipoles at the same latitude and longitude, followed by converging and shear motions. This temporal order of processes frequently invoked in magnetohydrodynamic models - emergence, converging motions, and shear motions - is critical for the storage of magnetic energy preceding strong solar eruptions that, under the right timing, location and direction conditions, can trigger severe space weather events at Earth.
\end{abstract}


\keywords{Sun: photosphere --- Sun: ARs photospheric dynamics --- Sun: flares --- Sun: geomagnetic effects}

%##########################################################################

\section{Introduction}

The solar atmosphere is an inherently dynamic system where magnetic fields play a crucial role in the manifestation of various solar phenomena. Among these, active regions (ARs) are particularly notable for their intense magnetic activity, which can lead to solar flares and coronal mass ejections (CMEs). These events are significant not only for their impact on space weather but also for their potential effects on Earth's magnetosphere.

ARs usually appear in narrow latitudinal belts between +35° and -35°, approaching the equator as the 11-year solar cycle advances. Preferred longitudes of sunspot formation, known as active longitudes, have been identified (e.g., \citealt{Bog82, Bal83}), although there are inconsistencies regarding their number, lifetime, location, and rotation rate \citep{Ber03}. Persistent active longitudes separated by about 180° have also been detected on different types of cool active stars (e.g., \citealt{Rod00, Ber02, Kor02}). 

The most powerful solar eruptions typically occur in ARs characterized by significant and unusual emerging magnetic flux, as observed, for instance, in AR 12673, which produced the most intense flare of Solar Cycle 24. In that case, the flux emergence rate reached an extraordinary peak value of approximately 1.12 $\times$ 10$^{21}$ Mx hr$^{-1}$. However, it is also important to consider that strong horizontal displacement of sunspots may play a crucial role in storing magnetic free energy (e.g., \citealt{Rom18}). \citet{Rom19} specifically demonstrated that shear motions observed along the polarity inversion line (PIL) of the main sunspots, when combined with intense magnetic fields exceeding 4000 G, can serve as a reference for studying not only the most powerful solar flares but also flares on other stars characterized by higher orders of magnitude.

One of the most famous events associated with a solar AR is the Carrington Event of March 1859, with a peak Dst-index reaching extraordinary value of $\sim$-589 nT, the most powerful geomagnetic storm on record (see \citealt{Cli13, Sai16, Bot19} and references therein). This AR was characterized by strong magnetic field concentrations and a particular compactness \citep{Car59}. Understanding the conditions and processes leading to such peculiar morphology and extreme events is essential for improving space weather prediction and mitigating its effects.

Recently, AR 13664, during its passage over the solar disc from May 2 to May 14, 2024, showed impressive activity with many recurrent high-intensity flares. Notably, two X-class flares occurred on May 9 and 10, each associated with a corresponding CME. These events triggered an extreme G5-level storm (K$_p$=9) on May 10-11, with geomagnetic indices, commonly used to assess the severity of a geomagnetic storm at Earth, attaining impressive values: solar wind speed exceeding $\sim$750 km s$^{-1}$ up to $\sim$1000 km s$^{-1}$ the following days, the north–south magnetic field component B$_z$ almost going down to –50 nT, and Dst-index about -412 nT around 03:00UT on 11$^{th}$ of May. AR 13664 presents a unique opportunity for studying the emergence and evolution of prolific ARs. Initially appearing at the East limb as a relatively ordinary AR — comprising fewer than 10 sunspots, each typically between 5,000 and 20,000 km in diameter, with a total area of less than 300 millionths of the solar hemisphere, a magnetic field strength between 1,000 and 3,000 G, and a relatively simple magnetic configuration, typically $\alpha$ or $\beta$, with few polarity inversions — it rapidly evolved within 3-4 days. The AR developed characteristics reminiscent of the one responsible for the Carrington Event, growing to over 100 sunspots often grouped in complex formations, with the largest spots exceeding 30,000 km in diameter, a total area approaching 1,000 millionths of the solar hemisphere ($\mu$Hem), magnetic field strengths surpassing 4,000 G, and a very complex $\beta\gamma\delta$ configuration with numerous close polarities.

This work focuses on the unusual magnetic flux emergence and peculiar horizontal motions during its development. These motions are investigated to understand their role in the rapid intensification and complexity of the magnetic field within the region. By analyzing high-resolution solar observation data, we aim to provide new insights into the mechanisms behind the formation of compact active regions and their potential for extreme solar activity.

In the next section, we describe the dataset used and the evolution of AR 13664 during its passage across the solar disc before the onset of the flares responsible for the G5 geomagnetic storm on May 10-11, 2024. The method of analysis and the results are described in Section 3, while the main conclusions are highlighted in Section 4.

%% #####################################################################
\section{Evolution of AR 13364}
%% #####################################################################

For this study, we utilized high-resolution data from {\it Helioseismic and Magnetic Imager} \citep[HMI;][]{Sch12} onboard the {\it Solar Dynamics Observatory} spacecraft \citep[SDO;][]{Pes12}. The HMI provides comprehensive observations of the solar photosphere, enabling detailed analysis of magnetic field configurations and motions within active regions.

We specifically used Space-weather HMI Active Region Patches (SHARPs) data products, based on vector magnetograms data series with a cadence of 12 minutes and a spatial resolution of 1 arcsecond per pixel \citep{Bob14}. These data were collected continuously from May 2, 2024, at 00:00 UT to May 10, 2024, at 23:48 UT, with an interruption on May 8 from 16:36 UT to 23:59 UT. These 1012 magnetograms allowed us to capture the entire passage of the AR, including the formation phase, the buildup of magnetic complexity, and the eruption phases associated with the X-class flares.

To gain a more comprehensive understanding of the photospheric features associated with AR 13664, we complemented the magnetogram data with continuum intensity images from HMI. These images provided valuable context for the magnetic field observations, aiding in the identification of sunspot structures and other photospheric phenomena associated with the active region's evolution.

%%%%%%%%%%%%%%%%%%%%%%%%%%%%%%%%%%%%%%%%%%%%%%%%%%%%%%%
\begin{figure}
\begin{center}
\includegraphics[trim=0 0 0 0, clip, scale=0.9]{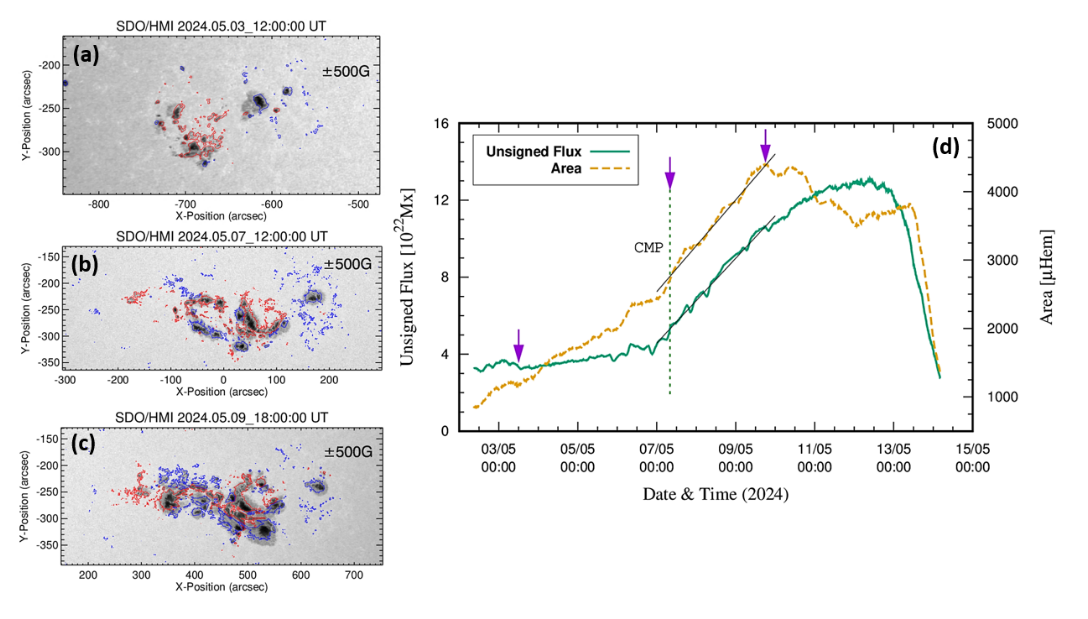}
\caption{In the left panels a sequence of three HMI continuum images describing the main evolution of the AR 13664 before the X8.7 flare occurrence has been reported. The blue and red contours correspond to -500 G and +500 G, respectively. The panel (d) displays the temporal evolution of the computed total unsigned magnetic flux (green) and the area in $\mu$Hem covered by the sunspots forming the AR (dark-yellow dashed line). The central meridian passage (CMP) is highlighted by vertical short-dashed line. The curves start at 18:00 UT on May 2, 2024. The three arrows indicate the times when the images reported in panels (a), (b) and (c) have been acquired. Best linear least-squares fits are drawn, depicting the significant growth phases (see text for details).}
\label{Fig1}
\end{center}
\end{figure}
%%%%%%%%%%%%%%%%%%%%%%%%%%%%%%%%%%%%%%%%%%%%%%%%%%%%%%%

%%%%%%%%%%%%%%%%%%%%%%%%%%%%%%%%%%%%%%%%%%%%%%%%%%%%%%%
%\begin{figure}
%\begin{center}
%\includegraphics[trim=40 150 0 240, clip, scale=0.76]{magnetic_flux_13664.ps}
%\caption{Magnetic flux vs. time of the AR NOAA 13664 during its passage over %the solar disc.}
%\label{Fig2}
%\end{center}
%\end{figure}
%%%%%%%%%%%%%%%%%%%%%%%%%%%%%%%%%%%%%%%%%%%%%%%%%%%%%%%

The global evolution of AR 13664 was marked by significant changes as it traversed the solar disk, culminating in a series of powerful X-class flares that triggered the geomagnetic storm on May 10, 2024. AR 13664 first appeared on the eastern limb of the Sun on May 1, 2024. Initially, the region was relatively small and magnetically simple (a $\beta$-class until late third of May, panel (a) of Fig. \ref{Fig1}). However, as it progressed across the solar disk, it rapidly developed in size and complexity, evolving into a $\beta\gamma\delta$-class (see \citet{Hal19} for a complete description of the complexity classes of ARs).

By May 5, 2024, the active region exhibited significant growth in both its magnetic field strength and area, with numerous sunspots forming and expanding (see panel (b) of Fig. \ref{Fig1}). This period was characterized by intense magnetic flux emergence, where new magnetic fields bubbled up from beneath the solar surface (see panel (d) of Fig. \ref{Fig1}). The region's complexity increased as these new fields interacted with the existing magnetic structures, leading to the formation of complex magnetic configurations with components significant proper motions and shearing episodes accompanied by the onset of a notable flux growth phase (see panels (c) and (d) of Fig. \ref{Fig1} and discussions in next paragraphs). 
Panel (d) of Fig. \ref{Fig1} illustrates the temporal evolution of the retrieved SHARP magnetic parameters `Unsigned Flux' and `Area'. Note that the area is calculated using the HMI line-of-sight magnetic field, while the unsigned flux is derived from the vector magnetic field components. Both parameters are based on the SHARP automated tracking system, which identifies magnetically strong-field structures within the AR (known as active pixels) after applying geometric corrections (refer to \citealt{Bob14, Hoe14, Bob15} for more details).

The rate of the total unsigned flux emergence increased dramatically just prior to the flares responsible for the severe geomagnetic disturbances. We quantified the growth rates for both area and flux by performing linear least-squares fitting over the period from May 7 to May 9. The resulting best-fit lines are shown in Fig. \ref{Fig1} (d). The flux exhibited a significant increase, transitioning from a slow growth phase (until mid May 6; 0.348 $\pm$0.001 $\times$ 10$^{22}$ Mx/day) to a rapid growth phase (starting around May 7th; 2.202 $\pm$0.012 $\times$ 10$^{22}$ Mx/day). A similar trend was observed for the area, although less pronounced compared to the flux, which increased from 353.47 $\pm$3.25 $\mu$Hem/day to a higher rate of 670.44 $\pm$5.26 $\mu$Hem/day. Interestingly, and in particular on May 9th  AR 13664 reached the peak of the area covered by its sunspots, i.e., $\sim$3494 $\mu$Hem, although the peak of the total unsigned magnetic flux ($\sim$13.53 $\times$ 10$^{22}$ Mx) has been reached about two days later.
Remarkably, this large magnetic flux rate in AR 13664 appears to be among the fastest flux emergence rates observed so far, exceeding even that of AR 12673 (August 2017) that hosted the most intense flare of SC24, and AR 12192 (October 2014) the largest AR in SC24 (refer to Figure 2 of \citealt{Sun24}).

%The rate of the total unsigned flux emergence has been estimated to be as high as $\sim$2 $\times$ 10$^{22}$ Mx per day from May 6 to May 9, 2024, just prior to the severe geomagnetic disturbances. In particular on May 9 AR 13664 reached the peak of the area covered by its sunspots, i.e., $\sim$3494 millionths of the solar hemisphere ($\mu$Hem), although the peak of the total unsigned magnetic flux ($\sim$13.53 $\times$ 10$^{22}$ Mx) has been reached about one day later. Interestingly, this large magnetic flux rate in AR 13664 appears to trace the fastest flux emergence observed in the modern solar observation era so far, exceeding even that of AR 12673 (August 2017) that hosted the most intense flare of SC24, and AR 12192 (October 2014) with its huge size-the biggest in SC24- (refer to Figure 2 of \citet{Sun24} ).

On May 9 AR 13664 showed pronounced shear motions indicative of significant magnetic stress building up within the region, usually a key precursor to solar flares. Indeed, the magnetogram analysis revealed strong horizontal flows and significant shearing along the PIL, where the positive and negative magnetic fields met and interacted. AR 13664 had become one of the most magnetically complex and active regions observed in the 25$^{th}$ solar cycle. We argue that, consequently, the magnetic field was most likely highly sheared, creating a highly unstable configuration prone to eruptions. This set the stage for the series of X-class flares. The first of these flares occurred at 9:13 UT on May 9, rapidly followed by several more intense flares within the subsequent 24 hours.

The flaring activity culminated in a major X3.98-class flare on May 10, 2024, which was directly responsible for the ensuing G5-class geomagnetic storm. The rapid release of magnetic energy during these flares was probably facilitated by the highly sheared magnetic fields that have been observed over the preceding days and that allowed to reach the enormous value of the flare index \citep{Li04, Rom07} of $\sim$5280\footnote{AR 13664 associated flare index is the highest for ARs of SCs 24 and 25 so far; AR 12673 the most active AR in SC24 had FI$\sim$3000.}, quantifying the flare productivity of the active region during its solar disk presence: 39 C-class, 61 M-class and 11 X-class flares. This sequence of events highlights the critical role of magnetic evolution and horizontal motions in driving extreme solar activity.

%% #####################################################################
\section{Results}
%% ###############################################################

The HMI observations were processed using standard SolarSoft routines to remove instrumental effects and correct for projection. We employed the Differential Affine Velocity Estimator for Vector Magnetograms \citep[DAVE4VM;][]{Sch05, Sch06} technique to analyze the horizontal motions of magnetic elements within the photosphere, using a window size of the apodizing window of 11 pixels (5$\arcsec$.5) which balances the need for capturing small-scale features while minimizing noise in the velocity estimations. The temporal cadence of the magnetograms was maintained at 12 minutes to ensure a high temporal resolution that captures the rapid evolution of magnetic structures. We used the standard configuration of DAVE4VM with default settings for the noise threshold and other algorithm-specific parameters, which have been optimized for HMI data in previous studies (e.g., \citet{Rom14}). This method provides a robust estimation of the velocity field from the time series of vector magnetograms, allowing us to quantify the shearing and converging motions critical to understanding the magnetic energy buildup and release processes in AR 13664 .

%%%%%%%%%%%%%%%%%%%%%%%%%%%%%%%%%%%%%%%%%%%%%%%%%%%%%%%
\begin{figure}
\begin{center}
\includegraphics[trim=30 80 0 140, clip, scale=0.56]{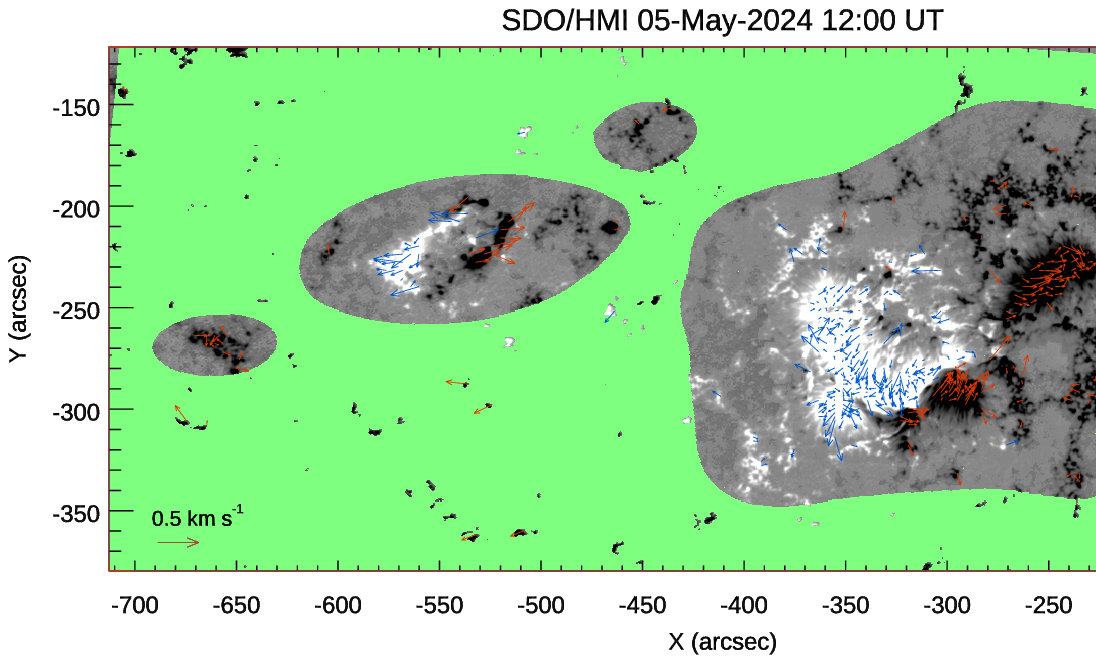}\\
\includegraphics[trim=30 80 0 140, clip, scale=0.56]{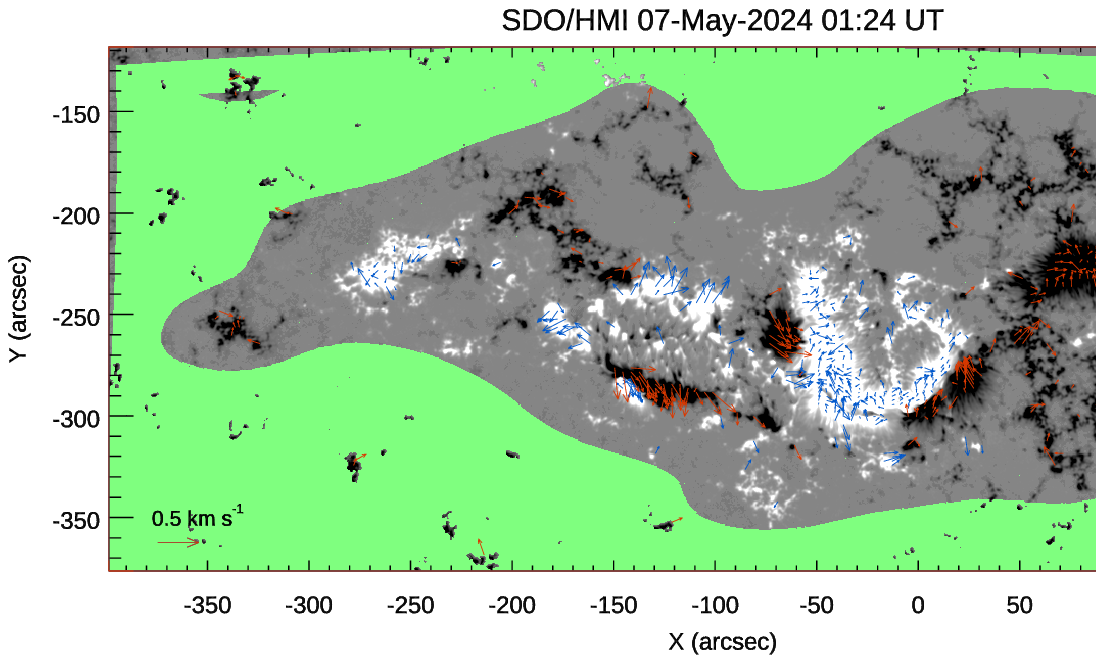}\\
\caption{Horizontal velocity maps obtained by DAVE4VM using SHARP HMI vector magnetograms, describing the phase of the maximum rate of magnetic flux emergence, from May 5 at 10:00 UT to May 7 at 10:00 UT. We used the color green to indicate region where no data are available in the SHARP magnetograms. A complete animation covering this phase from May 5 at 10:00 UT to May 7 at 10:00 UT is also available online. In the animation, the emergence of two main bipoles is observed on the eastern side of the preexisting magnetic features, with their opposite polarities gradually moving apart from each other.}
\label{Fig2}
\end{center}
\end{figure}
%%%%%%%%%%%%%%%%%%%%%%%%%%%%%%%%%%%%%%%%%%%%%%%%%%%%%%%

%%%%%%%%%%%%%%%%%%%%%%%%%%%%%%%%%%%%%%%%%%%%%%%%%%%%%%%
\begin{figure}
\begin{center}
\includegraphics[trim=30 80 0 200, clip, scale=0.56]{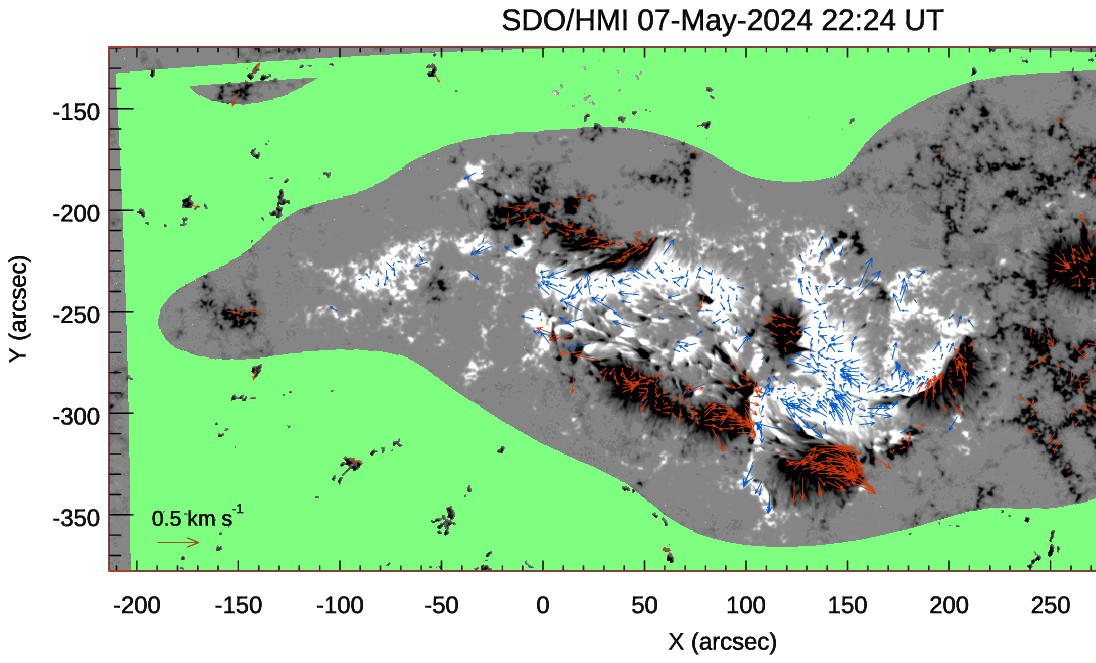}\\
\includegraphics[trim=30 80 0 140, clip, scale=0.56]{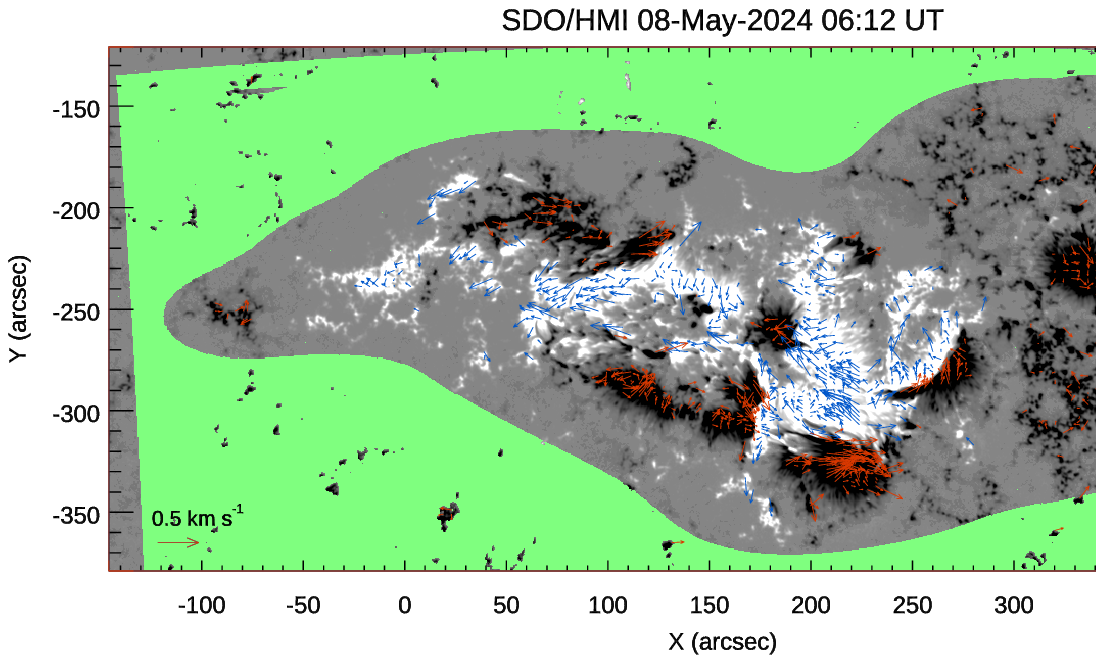}\\
\caption{Same of Figure \ref{Fig2} but describing the compaction of the AR 13664, from May 7 at 10:00 UT to May 8 at 10:00 UT. We used the color green to indicate region where no data are available in the SHARP magnetograms. A complete animation covering this phase from May 7 at 10:00 UT to May 8 at 10:00 UT is also available. In the animation, we observe a change in the direction of the previously emerged structures, which no longer move apart but begin to converge and compact.}
\label{Fig3}
\end{center}
\end{figure}
%%%%%%%%%%%%%%%%%%%%%%%%%%%%%%%%%%%%%%%%%%%%%%%%%%%%%%%

%%%%%%%%%%%%%%%%%%%%%%%%%%%%%%%%%%%%%%%%%%%%%%%%%%%%%%%
\begin{figure}
\begin{center}
\includegraphics[trim=30 80 0 200, clip, scale=0.56]{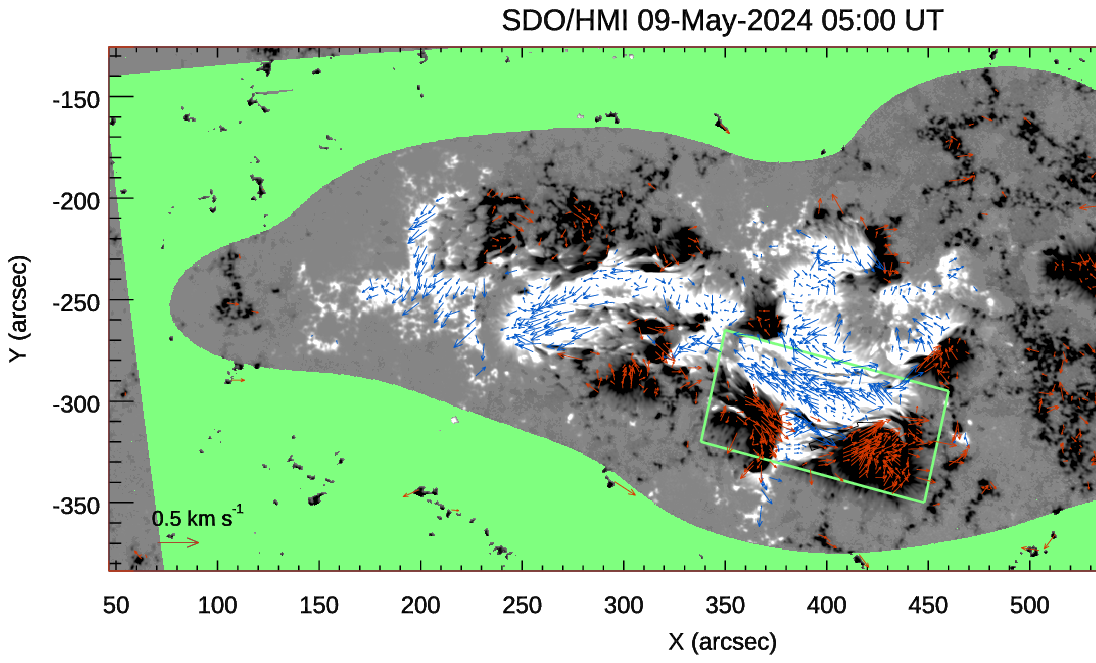}\\
\includegraphics[trim=30 80 0 140, clip, scale=0.56]{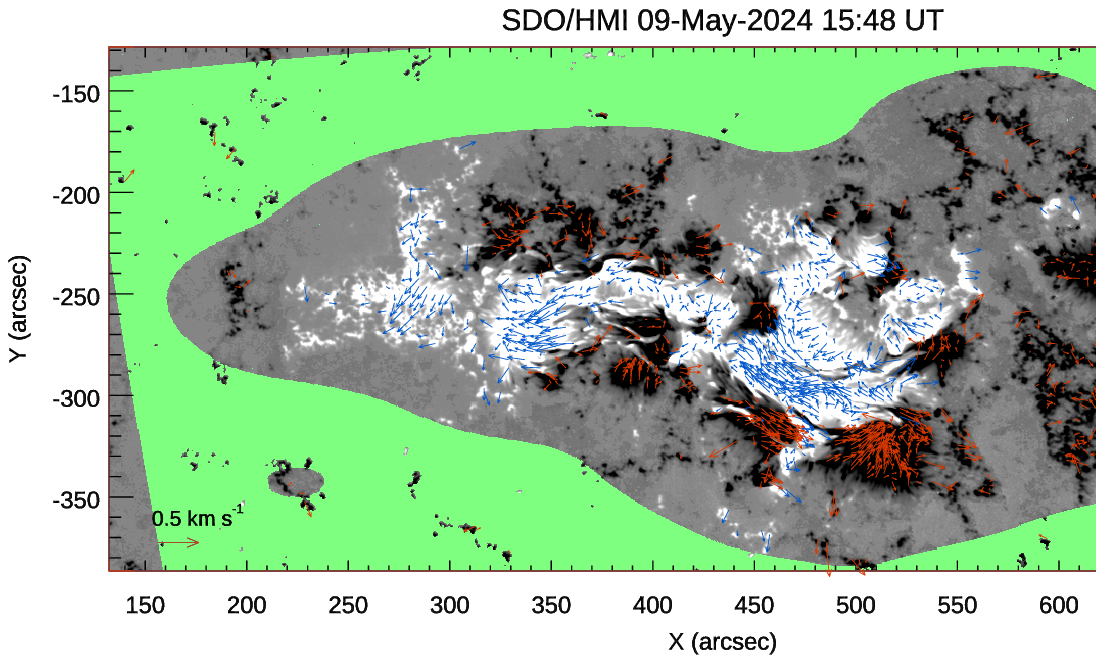}\\
\caption{Same of Figure \ref{Fig2} but describing the shearing motions, from May 8 at 10:00 UT to the end of the dataset. We used the color green to indicate region where no data are available in the SHARP magnetograms. The green box highlights the region where the shearing motions are located. A complete animation covering this phase from May 8 at 10:00 UT to May 10 at 24:00 UT is also available. In the animation, shearing motions can be observed along most of the main PIL of the AR, where the positive polarity predominantly moves toward the northeast, while the negative polarity moves toward the southwest. This pattern is consistently visible throughout the entire sequence.}
\label{Fig4}
\end{center}
\end{figure}
%%%%%%%%%%%%%%%%%%%%%%%%%%%%%%%%%%%%%%%%%%%%%%%%%%%%%%%

Our analysis identifies four distinct and key evolutionary stages in the AR's development: the early phase, the emergence phase, the compaction phase and the shear phase. Despite some overlap between these stages, we were able to clearly distinguish their sequential onset.

{\em The early phase.} AR 13664 appeared at the East limb at a latitude of 18°S as a common bipolar region, oriented along the East-West direction with the preceding negative polarity in the direction of the solar rotation. This phase, corresponding to the first 3-4 days of the AR passage across the solar disc as shown in Fig. \ref{Fig1}, does not show any significant evolution in terms of magnetic flux variations and area covered by the AR sunspots. Only the projection effects show some variation in the extent of AR 13664, while the magnetic flux remains just below 4 $\times$ 10$^{22}$ Mx, which is average for active regions at this phase of the solar cycle \citep{Sha23}.

{\em The emergence phase.} This phase started on May 5, when several magnetic bipoles began to emerge on the eastern side of the pre-existing sunspot system. The first new bipole appeared about 200$\arcsec$ to the east of the pre-existing field and was oriented along the East-West direction, with the negative polarity leading. In the horizontal velocity maps, this emergence is manifested by the progressive separation of the two polarities, with velocities of a few hundred meters per second oriented parallel to the equator (see the region at x=-500$\arcsec$, y=-220$\arcsec$ in the top panel of Fig. \ref{Fig2}). On May 6, a second bipole emerges between the initial AR location and the previously emerged bipole (see the region at x=-50$\arcsec$, y=-260$\arcsec$ in the bottom panel of Fig. \ref{Fig2}). This bipole, also characterized by a leading negative polarity in the direction of rotation, is initially oriented along the Southeast-Northwest direction. This new bipole appears to emerge with greater intensity compared to the previous one. This is evident not only in terms of the new magnetic flux that appears on the surface between May 6 and 7, but also from the separation velocities of its polarities, which reach up to 0.5 km s$^{-1}$. We note that the peak emergence rate of 2.202 $\pm$0.012 $\times$ 10$^{22}$ Mx/day was reached on May 7, although this phase continued in the following days, overlapping with the two subsequent phases. Before the AR reached the western limb, where projection effects become significant, the total unsigned magnetic flux was approximately $\sim$13.53 $\times$ 10$^{22}$ Mx.

{\em The compaction phase.} During this phase the emerging fluxes expanded horizontally, contributing to the overall growth of the active region. The horizontal velocity magnitudes in the southern part of the AR range from 0.5 to 1.0 km s$^{-1}$ (see the region at x=150$\arcsec$, y=-330$\arcsec$ in the top panel of Fig. \ref{Fig3}), consistent with the vigorous flux emergence seen in highly ARs \citep{Rom18}. In particular, we note that the combined process of new flux emergence and its westward propagation leads to the formation of a particularly compact active region. This region's morphological characteristics are reminiscent of the active region responsible for the Carrington Event. At the end of this phase, we observe a significant decrease in the separation velocity of the two polarities along the south-north direction. Additionally, a broad area characterized by a sea-serpent pattern of the magnetic field extends between them (see the bottom panel of Fig. \ref{Fig3}). These variations in the orientation of horizontal motions, which shifted from separating opposite polarities (as typically occurs during the emergence of individual flux tubes) to converging different magnetic structures within the AR, resulted in a reduction in sunspot fragmentation and the formation of exceptionally large sunspots, with the largest spots exceeding 30,000 km in diameter. Additionally, these compaction processes resulted in a decrease in the number of individual sunspots within the AR, from around 81 to approximately 43, despite the continued emergence of new magnetic flux.

{\em The shear phase.} Based on the horizontal velocity maps, we observe a correlation between magnetic flux complexity and intricate velocity patterns (Fig. \ref{Fig4}). Specifically, after the onset of the emergence and compaction phases, we identify strong shearing motions along the PIL, where the southern negative portion of the active region exhibits westward displacement, while the positive flux on the opposite side of the PIL moves eastward (see the green box around the region at x=400$\arcsec$, y=-300$\arcsec$ in the top panel of Fig. \ref{Fig4}). This shear persists for several hours, extending at least until the AR reaches the western limb, and affects the magnetic flux concentrations at the eastern edge of the AR (refer to the region at x=340$\arcsec$, y=250$\arcsec$ in the bottom panel of Fig. \ref{Fig4}). The peculiarity of these motions lies in their strength (up to 1.0 km s$^{-1}$), duration (over 2 days), and extent (about 200$\arcsec$ in latitude, i.e., from x=300$\arcsec$ to x=550$\arcsec$ in the bottom panel of Fig. \ref{Fig4}). These shear motions, probably driven also by differential rotation and convective flows, seems to be crucial in building up magnetic stress within the region, before the trigger of the main flares occurred in the AR. Alongside shearing, converging flows were also observed near the PIL. Specifically, the velocities of the two polarities of the second emerged bipole transitioned from diverging to converging (compare the arrows direction of the central part of the AR in the bottom panel of Fig. \ref{Fig2} with the bottom panel of Fig. \ref{Fig4}), thereby reducing the latitudinal extent of the AR in its central portion. These converging motions, reaching velocities up to 0.5 km s$^{-1}$, facilitated the accumulation of magnetic flux and likely intensified the magnetic field gradient.

%%%%%%%%%%%%%%%%%%%%%%%%%%%%%%%%%%%%%%%%%%%%%%%%%%%%%%%
\begin{figure}
\begin{center}
\includegraphics[trim=0 160 0 420, clip, scale=0.78]{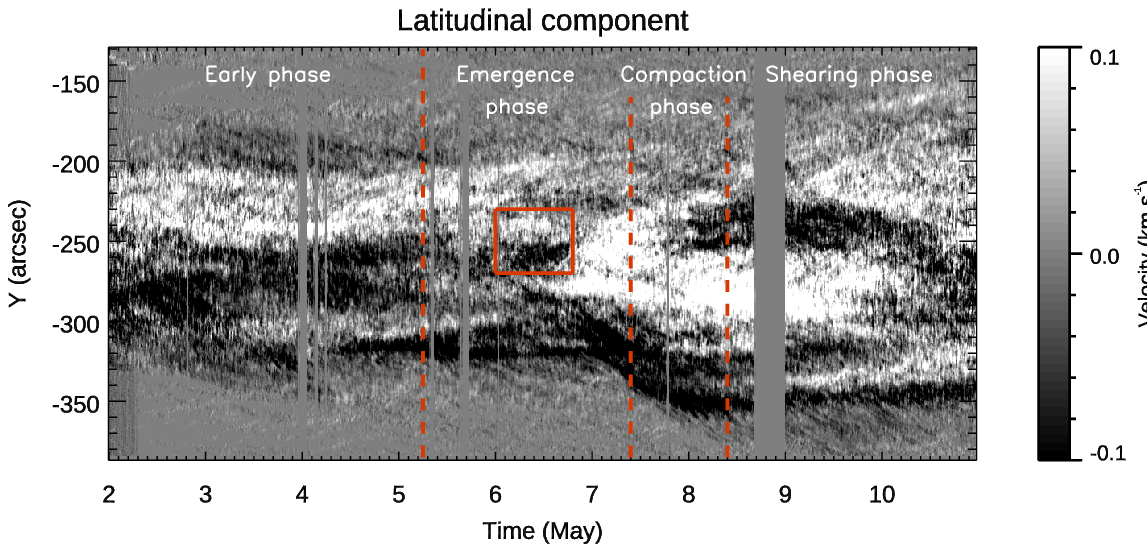}\\
\includegraphics[trim=0 120 0 420, clip, scale=0.78]{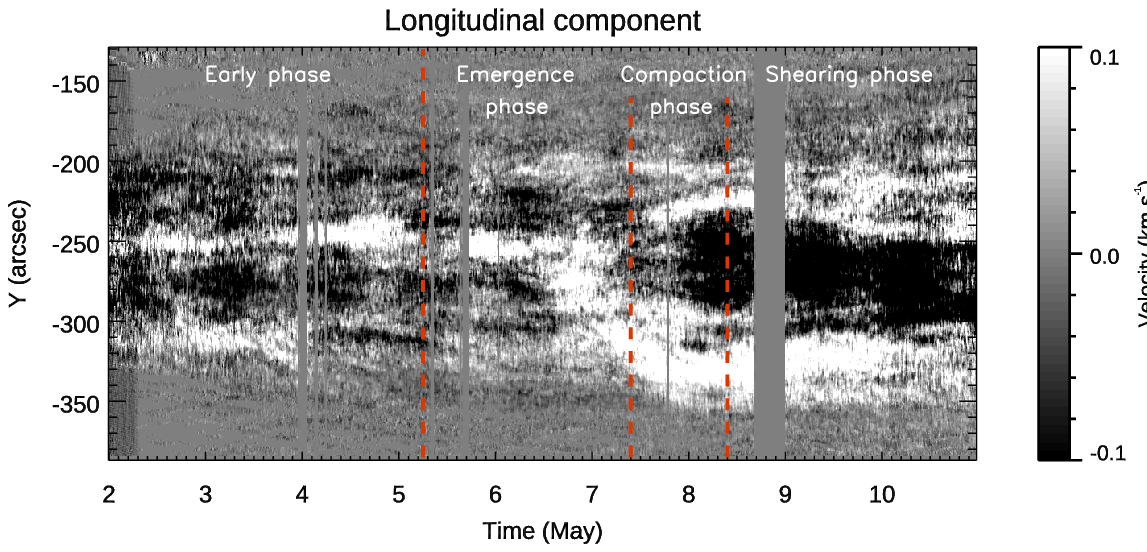}\\
\caption{Average of the latitudinal (top panel) and longitudinal (bottom panel) components of the horizontal velocities along the latitudinal extension of the AR over time. Positive (white) and negative (black) values correspond to northward and southward directions, respectively, for the latitudinal component. For the longitudinal component, positive (white) and negative (black) values represent westward and eastward directions, respectively. The red box highlights the emergence of the second main bipole. The gray vertical bars indicate periods of SHARP data gaps.}
\label{Fig5}
\end{center}
\end{figure}
%%%%%%%%%%%%%%%%%%%%%%%%%%%%%%%%%%%%%%%%%%%%%%%%%%%%%%%

Interestingly, the time evolution maps of the average latitudinal (top panel of Fig. \ref{Fig5}) and longitudinal (bottom panel of Fig. \ref{Fig5}) components of the horizontal velocities along the AR clearly confirm and emphasize the four phases described earlier. Initially, until May 5, the average velocities of both components do not exhibit significant variations. Specifically, in the average latitudinal components, there is a prevalence of northward and southward velocities above and below y=-250$\arcsec$, respectively, which can be interpreted as the initial expansion of the AR in latitude. The emergence phase becomes evident in both maps starting from 12:00 UT on May 5. The emergence of the second bipole between the initial AR location and the previously emerged bipole is characterized by significantly higher velocities reaching up to 0.5 km s$^{-1}$, which influence the average velocities with a much more intense signal compared to the previous phase. We note the effects of the emergence of this bipole in the prevalence of northward and southward directed velocities above and below y=-260$\arcsec$, respectively (see the red box in Fig. \ref{Fig5}). This corresponds to the location of the neutral line of the new magnetic system overlapping the pre-existing one. 

The converging phase becomes clearly evident starting from May 7, when the northward component at y=-250$\arcsec$ progressively shifts to a southward component in the average latitudinal velocity map. 

Given that the PIL is oriented longitudinally, the shear is noticeable in the strong eastward component above y=-300$\arcsec$, countered by a similarly significant westward component below it.

%% ###############################################################
\section{Discussion}
%% ###############################################################
Our detailed analysis of AR 13664, spanning from its initial appearance on May 1, 2024, to its peak activity on May 10, 2024, reveals a complex evolutionary path marked by substantial quantitative changes. Initially classified as a $\beta$-class region with a total unsigned magnetic flux of just below 4 $\times$ 10$^{22}$ Mx, AR 13664 rapidly evolved in size and complexity. By May 5, the active region expanded significantly, reaching a magnetic flux growth rate of 2.202 $\pm$0.012 $\times$ 10$^{22}$ Mx/day, a notable increase from the previous slower growth rate of 0.348 $\pm$0.001 $\times$ 10$^{22}$ Mx/day.

During the emergence phase, the total unsigned magnetic flux increased to approximately 13.53 $\times$ 10$^{22}$ Mx before the region reached the western limb. In comparison, historical reconstructions suggest that the active region responsible for the Carrington Event in 1859 had a total unsigned magnetic flux of around $5 \times 10^{22}$ Mx, although this estimate carries significant uncertainties due to the limitations of data from that period (see \citealt{Cli13}).

The compaction phase saw horizontal velocity magnitudes ranging from 0.5 to 1.0 km/s in the southern part of the region. This phase resulted in a reduction in the number of sunspots from around 81 to approximately 43, despite ongoing flux emergence. The largest sunspots in AR 13664 exceeded 30,000 km in diameter, whereas historical records indicate that the Carrington AR had sunspots possibly larger than 50,000 km \citep{Hay19}.

The shear phase, marked by strong shearing motions along the PIL, exhibited velocities up to 1.0 km/s. This shearing persisted for over 2 days and covered a latitudinal extent of approximately 200$\arcsec$. The analysis of horizontal velocities confirmed these phases with detailed maps showing shifts from diverging to converging motions and significant increases in the latitudinal and longitudinal components of velocity.

The culmination of these processes led to a series of intense X-class flares, with the flare index reaching approximately 5280, the highest recorded for ARs in solar cycles 24 and 25. This intense activity ultimately triggered a major geomagnetic storm on May 10, 2024. For context, the Carrington Event produced the most intense geomagnetic storm on record, with estimated Dst indices exceeding -1600 nT, which is several times stronger than typical modern geomagnetic storms \citep{Sis06}. This comparison underscores the critical role of magnetic flux emergence, shearing, and horizontal motions in driving extreme solar activity and its impact on the solar and geomagnetic environment.

AR 13664 exhibited notable similarities and differences when compared to the active region responsible for the Carrington Event of 1859. Both regions were characterized by a compact and highly complex magnetic configuration, with strong magnetic fields and multiple sunspots, indicative of intense magnetic activity. This compactness and complexity were key features that contributed to their significant flare activity, as both regions unleashed multiple X-class flares.

However, despite these similarities, there were also important differences between the two regions. AR 13664, while exhibiting intense activity, was smaller in size compared to the Carrington Event active region, which was notably larger and more extensive (see \citealt{Hod59}). This difference in scale had significant implications for the geomagnetic impact of each region. The geomagnetic storm induced by AR 13664, while significant and classified as a G5 storm, was less intense than the Carrington Event, which produced a superstorm with extreme global geomagnetic disturbances (see \citealt{Tsu03, Cli13}).

Additionally, the duration of active phases between AR 13664 and the Carrington AR also showed notable differences. The emergence phase of AR 13664, characterized by a rapid increase in flux, spanned approximately 4 days, whereas the emergence phase of the Carrington AR, based on historical records, was more abrupt and occurred over a shorter timescale. The shear phase in AR 13664, marked by significant shearing motions along the PIL, persisted for over 2 days, covering a latitudinal extent of approximately 200". The Carrington AR, most apparently, exhibited a shorter but more intense shearing phase, which played a crucial role in the rapid release of magnetic energy, ultimately leading to the subsequent extreme geomagnetic storm \citep{Hay19}).

These differences in the duration and nature of the active phases highlight the variability in the lifecycle of solar active regions, even among those capable of producing extreme space weather events. While AR 13664 exhibited a more prolonged and gradual buildup of activity, the Carrington AR's rapid and intense phase transitions were key factors in its historical significance.

%% ###############################################################
\section{Conclusions}
%% ###############################################################

Our investigation of AR 13664 offers valuable and key insights into the characteristics of solar ARs, particularly when compared to other super ARs documented in the literature. On the one hand, the peak of the unsigned magnetic flux reached by AR 13664, of approximately 13.53 $\times$ 10$^{22}$ Mx, is comparable to other super ARs such as AR 12192, which exhibited a peak flux of about 2 $\times$ 10$^{23}$ Mx \citep{Sun15}. On the other hand, and in terms of flux emergence rate, AR 13664 had a peak rate significantly higher than that observed in some other large ARs, such as AR 12192 (peak rate of approximately 3.4 $\times$ 10$^{22}$ Mx/day) or AR 12673 (peak rate of approximately 2.9 $\times$ 10$^{22}$ Mx/day) \citep{Sun17}.

The largest sunspots in AR 13664 had diameters of up to 30,000 km. For comparison, the largest sunspots observed in AR 12192 were about 40,000 km in diameter \citep{Jai17}. Instead, during the compaction phase the number of sunspots in AR 13664 decreased from about 81 to 43, within the range reported for largest super ARs observed since 1996 up to now, such as AR 12192, AR 10486, AR 9393, where the number of sunspots varied between 60 and 108 during their peak activity.

We believe that the identified sequence and order of evolutionary phases in this AR are likely fundamental to the formation and development of such complex solar ARs. Indeed, the subsequent emergence of magnetic field concentrations one after another at the same latitude and, importantly, at the same longitude, followed by predominant converging motions among the various photospheric magnetic structures, enables the strengthening of the magnetic field and the concentration of alternating magnetic polarities. Only subsequently vigorous shear motions along the PIL contribute significantly to energy buildup within the active region. These motions facilitate the accumulation and redistribution of magnetic flux, leading to the development of complex magnetic configurations. On the contrary, a different sequence/order in these phases could result in an earlier release of accumulated energy and the consequent diffusion of the magnetic field, preventing the formation of a compact AR like the AR 13664 and the one responsible for the Carrington event. This result is further validated by the similar temporal evolution of the total magnetic flux observed in other super ARs as reported in the literature (e.g., \citealt{Rom07, Smy10}).

We also remark that the driving mechanisms behind the observed horizontal motions in AR 13664 could be linked to differential rotation, convective flows, and magnetic instabilities in the solar photosphere. Differential rotation could be one of the causes that contribute to shear motions along the PIL, while convective motions could drive converging flows and facilitate the transport of magnetic flux. The interaction of emerging magnetic fields with pre-existing structures also could play a significant role in shaping the dynamics of solar active regions. In particular, in AR 13664, we observe the appearance of two subsequent bipoles, corresponding to additional flux tubes at the same latitude and longitude as the pre-existing field. This behavior can be attributed to the higher rate of magnetic buoyancy and emergence during the solar cycle's maximum phases. When preferred longitudes of sunspot formation also coincide in latitude, we observe this succession of recurrent, intense flares attributable to the interaction between emerging and pre-existing magnetic fields \citep{Zha07}. This interaction is accompanied by a significant accumulation of energy due to the field's intensity and shear. 
These processes are essential for understanding the initiation of solar flares and CMEs from active regions like AR 13664, as well as similar activity observed in young sun-like stars (e.g., \citealt{Lan09}). Our study specifically highlights how horizontal motions and shearing contribute to the magnetic evolution of solar active regions, which in turn influences their potential to generate significant space weather events. By deepening our understanding of these dynamics, we can enhance predictive models for both solar and stellar activity, thereby improving our ability to anticipate and mitigate the impacts of extreme events on Earth's technological systems.

%% #############################
\acknowledgments
%% ############################# 
 We are thankful to the editorial board and anonymous reviewers for their valuable and constructive suggestions, that have significantly enhanced this paper. This work was supported by INAF (Bando per il finanziamento della Ricerca Fondamentale 2022 - Study of the correlation between the solar activity and the geomagnetically induced currents in gas pipelines systems- and Bando per il finanziamento della Ricerca Fondamentale 2023 - IDEA-SW project), by ASI under contract with INAF no. 2021-12-HH.0 “Missione Solar-C EUVST – Supporto scientifico di Fase B/C/D” and no. 2022-29-HH.0 "MUSE".
 The research work of A. Elmhamdi in this project was supported by King Saud University’s Deanship of Scientific Research and College of Science Research Center in Saudi Arabia.\\
We aknowledge the use of the different facilities, databases and tools appearing in our paper: SDO (AIA, HMI).\\


%% in parentheses, after the keyword, but they will not be verified.

%{\it Facilities:} \facility{SDO (AIA, HMI), SOHO (LASCO), INAF-OACT, GONG , e-callisto, RSTN network}.



%\appendix

%\section{Appendix material}

%%% ###################################

\clearpage

%%%%%%%%%%%%%%%%%%%%%%%%%%%%%%%%%%%%%%%%%%%%%%%%%%%%%%%
%\begin{figure}
%\begin{center}
%\includegraphics[trim=0 0 0 0, clip, scale=0.64]{Fig1_Global.eps}
%\caption{Panel (a): colorized magnetogram full-disc Sun (24th Dec, 2023 at 12:00UT). The main ARs presented on the solar photosphere are reported. Positive or negative polarity sunspots and other strong field regions appear blue or red with dark umbrae. Panel (b): time profiles of the GOES X-ray flux (upper curve: 1-8 \AA; lower curve: 0.5-4 \AA{}). The arrows point to the peak times of the three M-class flares of day 24 Dec. Panels (c) ~\& ~(d): the  first two CMEs of December 24th (see text for details).}
%\label{Fig1A}
%\end{center}
%\end{figure}
%%%%%%%%%%%%%%%%%%%%%%%%%%%%%%%%%%%%%%%%%%%%%%%%%%%%%%%
%%%%%%%%%%%%%%%%%%%%%%%%%%%%%%%%%%%%%%%%%%%%%%%%%%%%%%%
%\begin{figure}
%\begin{center}
%\includegraphics[trim=0 0 0 0, clip, scale=0.6]{Fig2_FilamEvol5.eps}
%\caption{Cutout images in three different wavelengths (H$\alpha$, AIA 304\AA{} and AIA 193\AA). Width and height are respectively 1020 ~\&~1490 arcsec. The most important characteristics specifically related to the development of the filaments (labeled “Fi1” and “Fi2” on a.1. image) indicated (refer to text for more details).}
%\label{Fig2A}
%\end{center}
%\end{figure}
%%%%%%%%%%%%%%%%%%%%%%%%%%%%%%%%%%%%%%%%%%%%%%%%%%%%%%%

%%%%%%%%%%%%%%%%%%%%%%%%%%%%%%%%%%%%%%%%%%%%%%%%%%%%%%%
%%%%%%%%%%%%%%%%%%%%%%%%%%%%%%%%%%%%%%%%%%%%%%%%%%%%%%%
\end{document}